\documentstyle[aps,epsf]{revtex}

\def\r{{\bf r}}

\def\q{{\bf q}}
\def\B{{\cal B}}
\def\vo{v_0}
\def\al #1{\frac{a^#1}{\lambda^#1}}
\def\all{\frac{a}{\lambda}}
\def\alls{a/\lambda}
\def\als #1{a^#1/\lambda^#1}
\def\sqpi{\sqrt{2\pi}}
\def\He#1{${}^{#1}{\rm He}$}
\def\erf{\mathop{\rm erf}\nolimits}
\def\const{{\rm const}}
\def\Sp{\mathop{\rm Sp}\nolimits}

\def\be{\begin{eqnarray*}}
\def\ee{\end{eqnarray*}}
\def\be{\begin{eqnarray}}
\def\ee{\end{eqnarray}}

\begin{document}

\title{
Thermodynamics of the Bose-System with a Small Number of Particles
}

\author{I.~O.~Vakarchuk\footnote{\tt chair@ktf.franko.lviv.ua}, 
A.~A.~Rovenchak\footnote{\tt andrij@ktf.franko.lviv.ua}
\\
Department for Theoretical Physics of Lviv National University,\\
12~Draghomanov St., UA--79005, Lviv, Ukraine}


\maketitle

\begin{abstract}
A theoretical description of the interacting Bose-system is proposed.
It is based on the extrapolation of the results obtained for the systems
with a small number of particles $N=2,\ 3,\ 4$, etc. to the bulk case of
$N=\infty$. It is shown that already the system with $N=12,\ 13$ behaves
almost as a bulk in a wide temperature range. Special attention is paid
to the phase transition in these systems. The hard sphere potential is used
in calculations. The sequence of heat capacity maxima is approximated as 
$C^{\rm max}_N/N\simeq13.6-aN^{-\varepsilon}$ with $\varepsilon=0.0608$
giving the value of bulk heat capacity as 13.6 while experimental
value is close to 16. The temperature of $\lambda$-transition is
estimated as 2.1--2.3~K versus experimental 2.17~K. 
Quite good qualitative and satisfactory 
quantitative agreement with the experimental data has been achieved.

\end{abstract}

\section{Introduction}
Liquid \He4 is the only atomic system having its properties determined
mainly by quantum effects and the most essential among the latter
is the statistical effect of identical particles. 
Due to it the macroscopic number of the \He4 atoms can 
have a zero momentum at some conditions, i.~e., it can 
occupy a single quantum state.
The phenomenon of bosons accumulation at the lowest energy level
is called Bose-Einstein condensation (1924). According to London's assumption
(1938), it is believed that superfluidity, 
$\lambda$-transition, and other unusual properties of liquid \He4
are a consequence of this very effect.

However, despite the lapse of time that has passed since the discovery
the phenomena of $\lambda$-transition by Keesom and Clausius in 1932
and the superfluidity by Kapitsa in 1938
until now noone has managed to create a comprehensive microscopic theory 
explaining the properties of 
strongly interacting Bose-systems such as \He4 within the whole temperature 
range of $0< T<\infty$.

Bogoliubov\cite{Bogoliubov} was the first to show how one can build 
a theory
of real helium proceeding from the approximate second quantization techniques for
the model of weakly-imperfect Bose-gas.
This theory works well for $T\lesssim 1$~K, especially while $T\to 0$.

The subsequent theoretical studies brought about the appearance of the quantitative
theory of the ground state of a strongly-imperfect many-boson system
leading to a good agreement with the experiment and the creation of the
theoretical description of low-energy excitations which are responsible for
the low-temperature behaviour of thermodynamic functions.
The experimental investigations of the spatial structure by means of X-ray 
and
neutron diffraction and of the structure of the energy spectrum and the condensate
fraction by means of the inelastic neutron scattering revealed a fairly 
good theoretical explanation.

As a consequence of the experimental evidences
it was supposed for a long time that \He4 heat capacity had
a logarithmic divergency at the transition temperature 
$T_{\lambda}$.
Brout showed in his early works~\cite{Brout} based
on the Hartree--Fock method that it was impossible to find an explanation
of the observed heat capacity behaviour within simple theoretical approaches.
There is an interesting notice regarding this question
Feynman made in his book~\cite{FeynmanSM}:
``The explanation of this behavior
is left as an exercise for the reader. If successful, publish!''

Feynman suggested that the $\lambda$-transition should be studied
using the
full $N$-particle density matrix for the ideal Bose-gas with a 
phenomenologically introduced effective mass for particles~\cite{Feynman53}.
This approach leads only to a shift of the Bose-condensation temperature
in comparison with the ideal case
leaving the form of the heat capacity curve
unchanged.

In the early 1970s the studies of the $\lambda$-transition
were started in the renormalization group (RG) approach for the theory
with the degenerate two-component order parameter.
The rigorous grounds of this approach based on the functional integration 
techniques are given in Refs.~\ref{Langer68},~\ref{Vak:coh_st}. They use the coherent states 
representation with the complex order parameter the absolute value of which
equals $\sqrt{N_0}$ where $N_0$ is the number of atoms in Bose-condensate.

In this approach the heat capacity critical exponent 
was calculated as $\alpha=1/8$.
It obviously does not correspond to the logarithmic divergence
for which $\alpha$ equals zero.
However, taking into account that the calculated value of $\alpha$
is a small magnitude, the experimental confirmation of this result required
very precise measurements which became available only recently~\cite{Lipa96}. 
Before it, there was even no certainty about the sign of $\alpha$.
The measurements showed that the critical exponent $\alpha$ is negative, i.~e.,
the heat capacity is finite at the transition temperature.

Recently the theoretical confirmation of this result appeared.
Kleinert~\cite{Kleinert98,Kleinert99} obtained for $\alpha$ the respective
values of $\alpha=-0.01294\pm0.00060$ and $\alpha=-0.0120\pm0.0009$, 
while the experiment gives $\alpha=-0.01285\pm0.00038$~\cite{Lipa96}.
One can find a bit different approach to the critical exponents calculation
in the paper by Campostrini {\it et al}~\cite{Campostrini00}.
The authors obtained the value of $\alpha=-0.0150\pm0.0017$. 
But, despite a very good agreement of these results with the experiment,
RG works only in the immediate vicinity of the transition point $T_{\lambda}$
and gives no possibility of getting some information of the 
thermodynamic functions behaviour in a wider near-critical range,
e.~g., for $|T-T_{\lambda}|\sim 10^{-3}$~K.

In addition to these theoretical investigations, the numerical studies
of \He4 thermodynamic properties were conducted.
The Diffusion Monte Carlo (DMC) techniques are applied for the ground state
calculations, and the Path Integral Monte Carlo (PIMC) is used for
finite temperatures.
These techniques have been applied since the 1960s.
At present, PIMC is the most effective method.
Its only essential imperfection lies in the fact that it is impossible
to use it for the ground state ($T=0$~K). But one can avoid this problem
by considering the limit of the results at $T\to0$~K. 
A detailed analysis of PIMC might be found in the work by
Ceperley~\cite{Ceperley95}. It appears that the numerical approach
provides a good agreement with the experimental data in a wide temperature
range on the whole, except for the vicinity of the $\lambda$-transition point.

Thus, we have the situation when no satisfactory theoretical description 
of \He4 exists for a considerable temperature range of 
$1\ {\rm K}\lesssim T<T_{\lambda}$ and $T>T_{\lambda}$, 
and the only way out is the application of numerical techniques.
It gives some space for the investigations in this domain.
Not claiming the solution of the problem, we still propose here
the model which, in our opinion, can near us to the understanding 
of the processes
in many-boson system via the consideration of systems with a small number
of particles.

For the latter one can expect a much more precise description in terms
of statistical mechanics in comparison with what might be reached
while directly considering a many-particle system.

The aim of the present work consists in obtaining a thermodynamic description
of an interacting many-boson system basing on the results for the systems with
a small number of particles $N$. 
We start our analysis from $N=2,\ 3,\ 4$, etc. in order to extrapolate
these results for the case of $N=\infty$ after establishing some
dependencies on $N$ and the temperature.
We pay a special attention to the fact that even the system consisting of
a small number of particles exhibits the peculiarities on the heat capacity
curve such as the maximum which means the phase transition in the limit
of $N\to\infty$.
We will show at the same time that the system with $N=12,\, 13$
tends to reveal qualitatively an ``almost bulk'' behaviour in a wide 
temperature range.

The boson systems with a relatively small number of particles
have been experimentally obtained only in the last couple of years 
as particles in traps
(see Refs.~12,~13 and references therein). 
It gives us a possibility of developing some methods for such systems and
a further comparison of the results with the experimental data.

We use the partition function formalism to calculate
thermodynamic characteristics of the free Bose-system.
A method somewhat similar to this one was used in 
works~\cite{Deng,Schmidt98} for the
study of the trapped finite Bose-systems: the recursion formulae
for the partition function proposed by P.~Borrmann and 
G.~Franke~\cite{Borrmann}. But our expressions do not take the restriction
for the system energy to be written as a sum of one-particle energies. So,
one can use them not only for the ideal gas but for the interacting system 
as well.

The paper is organized as follows.
In Section~II some known initiating expressions for 
the calculating the thermodynamic function are given in short. 
In Sections~III and~IV we consider how the proposed method works
in the case of an ideal Bose-gas the results for which are well-known
in the thermodynamic limit when the number of particles $N\to\infty$, 
The next step is the determination of the asymptotic behaviour
of our results depending on the number of particles $N$.
We will try to expand the found regularities for the case of interacting 
bosons.
The corresponding calculations for the hard-sphere potential are given
in Section~V.

Generally, it is impossible to obtain a non-analyticity point of 
the heat capacity function having finite number of particles.
Therefore, no ``pure'' phase transition can be detected in this model.
But we expect a certain qualitative agreement with the experiment
to be quite sufficient for the approximation used.

\section{General statements}
We consider the system of $N$ particles with the Hamiltonian $\hat H_N$.
The partition function $Z_N=\Sp e^{-\beta\hat H_N}$
where $\beta$ is inverse temperature, $\beta=1/T$.

In the coordinate representation we have
\be
Z_N=\int d\r_1\dots \int d\r_N \hskip 4pt R_N(\r_1,\dots,\r_N \hskip 2pt \vert
\hskip 2pt\r_1,\dots,\r_N) ,
\ee
where $R_N$ is the density matrix. The particles are bosons of 
the mass $m$ (\He4 atoms)
with their coordinates $\r_i$ restricted in volume $V$.
In this case $R_N$ is given by \cite{JPS97}
\be
R_N(\r_1,\dots,\r_N \hskip 2pt \vert
\hskip 2pt\r'_1,\dots,\r'_N)=P_N(\r_1,\dots,\r_N) \,
\frac1{N!}\frac1{\lambda^{3N}}
\sum_Q
\exp\left( -\frac{\pi}{\lambda^2}\sum_{j=1}^N(\r_j-\r'_{Q_j})^2\right),
\ee
where $Q$ is counting permutations of $(1, \dots\,, N)$ and
$\lambda$ is the thermal de Broglie wavelength
\be
\lambda=\left( {2\pi\beta \hbar^2} \over {m} \right)^{1/2}.
\ee
The factor $P_N$ takes into account the interaction and will be described
further.

The free energy $F_N$ is given by
\be
F_N=-\frac1{\beta}\ln Z_N,
\ee
total energy
\be
E_N=\frac{\partial}{\partial\beta}(\beta F_N),
\ee
and heat capacity
\be \label{CNdef}
C_N=\beta^2 \left\lbrack \frac{Z_N^{\prime\prime}} {Z_N}
- \left( \frac{Z_N^{\prime}} {Z_N} \right)^2 \right\rbrack,
\ee
the primes mean derivation over $\beta$.

\section{Partition function for the system of free particles}
The ideal Bose-gas is a well-studied system for which the expressions
for the thermodynamic functions are known. Thus, it can be a good test
example for the applicability of our method and we are going to use this 
fact here.

On calculating the thermodynamic functions of the ideal Bose-system
with a small number of particles and comparing them with the well-known
results for infinite system one can detect some regularities showing
the point when several particles start to demonstrate the
statistical behaviour.

One can find an expression for the partition function of $N$ ideal bosons
in the explicit form. For this purpose we write the density matrix $R_N$
as follows:
\be
R_N(\r_1,\dots,\r_N \hskip 2pt \vert
\hskip2pt\r_1,\dots,\r_N)=\frac1{N!}\frac1{\lambda^{3N}}\Delta_N,
\ee
where
\be\label{Delta_def}
\Delta_N=\left|
\begin{array}{ccccc}
1 & K_{12} & K_{13} & \ldots & K_{1N} \\
K_{21} & 1 & K_{23} & \ldots & K_{2N} \\
\vdots &   & \ddots &        &        \\
K_{N1} & K_{N2} & K_{N3} & \ldots & 1 \\
\end{array}
\right|_{+},
\ee
and
\be
K_{ij}&\equiv& K(r_{ij})=e^{-\pi r_{ij}^2 / \lambda^2}
,\qquad r_{ij}=|\r_{ij}|,
\qquad \r_{ij}=\r_i-\r_j.
\ee

The subscript ``$+$'' means that every ``$-$'' in the expression for
the determinant should be substituted with ``$+$''.

The partition function for free particles is
\be\label{Z_int}
Z_N=\frac1{N!}\frac1{\lambda^{3N}}\int d\r_1\dots\int d\r_N \hskip3pt\Delta_N.
\ee

The ``determinant'' (\ref{Delta_def})
might be written as \cite{JPS97}
\be \label{Delta_expan}
\Delta_N&=&\Delta_{N-1}^{(1)}+\sum_{1<i\le N}
K_{1i}K_{i1}\Delta_{N-2}^{(1,j)}
+\sum_{1<i<j\le N}
(K_{1i}K_{ij}K_{j1}+K_{1j}K_{ji}K_{i1}) \Delta_{N-3}^{(1,i,j)}+\dots,
\ee

The superscripts in $\Delta_{N-l}$ mean the lack
of the correspondent rows and columns in the expression for $\Delta_{N-l}$. 

Integrating (\ref{Z_int}) we obtain
\be
Z_N&=&\frac{1}{N}\left\{\frac{V}{\lambda^3} Z_{N-1}+
\frac{1}{(\lambda^3)^2}Z_{N-2} \int d\r_1 \int d\r_2
\hskip3pt K_{12} K_{21}
+\frac{1}{(\lambda^3)^3} Z_{N-3}
\int d\r_1 \int d\r_2 \int d\r_3 \hskip3pt K_{12}K_{23}K_{31} + \dots
\right\},
\ee

After the Fourier transformation
\be
K(r)&=&\frac 1 V\sum_{\q}e^{i\q\r}K_{\q},\\
K_{\q}&=&\int d\r\hskip3pt e^{-i\q\r}K(r)=\lambda^3 e^{-\lambda^2 q^2/4\pi}
\ee
the partition function is
\be \label{ZNid}
Z_N=\frac1{\rho\lambda^3}\sum_{l=1}^N\frac{Z_{N-l}}{l^{3/2}},
\qquad Z_0\equiv1.
\ee

One can use this expression to calculate the partition function
of the system. We will make such calculations in the next section.

\section{Main results for the free particles}
It is possible to calculate heat capacities $C_N$ for $N=1,\ 2,\ 3,\, \ldots$
particles sequentially using expressions (\ref{CNdef}) and (\ref{ZNid}).
In addition, we can show explicitly the low- and high-temperature
asymptotics for $C_N$, namely:
\be
&&\left.\frac{C_N}N\right\vert_{T\to\infty}=\frac32,\\
&&\left.\frac{C_N}N\right\vert_{T\to 0}=\frac32\frac1N+A_N T^{3/2},
\qquad A_N=\const {\rm\ with\ respect\ to\ }T.
\ee

In the case of $N=\infty$ this temperature dependencies coincide with 
the known results for the ideal Bose-gas~\cite{Huang}. 
For low temperatures one has
\be
\left.\frac{C_N}N\right\vert_{N=\infty}=\frac{15}4\frac{\zeta(5/2)}
{\zeta(3/2)}\left(\frac T{T_c}\right)^{3/2};
\ee
the high-temperature limit being 3/2. The critical temperature 
$T_c$ (the non-analyticity point on the heat capacity curve)
in this case is defined from the equation
\be
\rho\lambda_c^3=\zeta(3/2),
\ee
where $\lambda_c$ is the thermal de Broglie wavelength for $T_c$, and
$\zeta$ is the Rieman $\zeta$-function, $\zeta(3/2)=2.612\dots\;$.
Taking $\rho=0.02185$~\AA$^{-3}$ for liquid \He4
we obtain $T_c=3.138$~K.

In this work we consider the systems with different $N$ from 2 to 1200.
As was mentioned above it is impossible to obtain the non-analiticity
in the heat capacity, so we take the point of $C_N$ maximum as
the critical temperature $T_c^{(N)}$.

It appears that those values could be described
by the linear dependence
\be \label{extr1}
\rho\left[\lambda_c^{(N)}\right]^3 \simeq 2.61 - \frac c{N^{1/3}},
\quad c\simeq 1.9
\ee
with $\lambda_c^{(N)}$ being thermal de Broglie wavelength for $T_c^{(N)}$.
One can easily notice that 2.61 is very close to $\zeta(3/2)=2.612\dots\;$.

We also have a similiar correlation for
the $C_N$ maxima $C_N^{\max}/N$.
So, one should expect that the limit $N\to\infty$ gives the correct
curve (see Figs.~1 and 2).

\bigskip
\begin{figure}
\epsfxsize=85mm
\centerline{\epsfbox{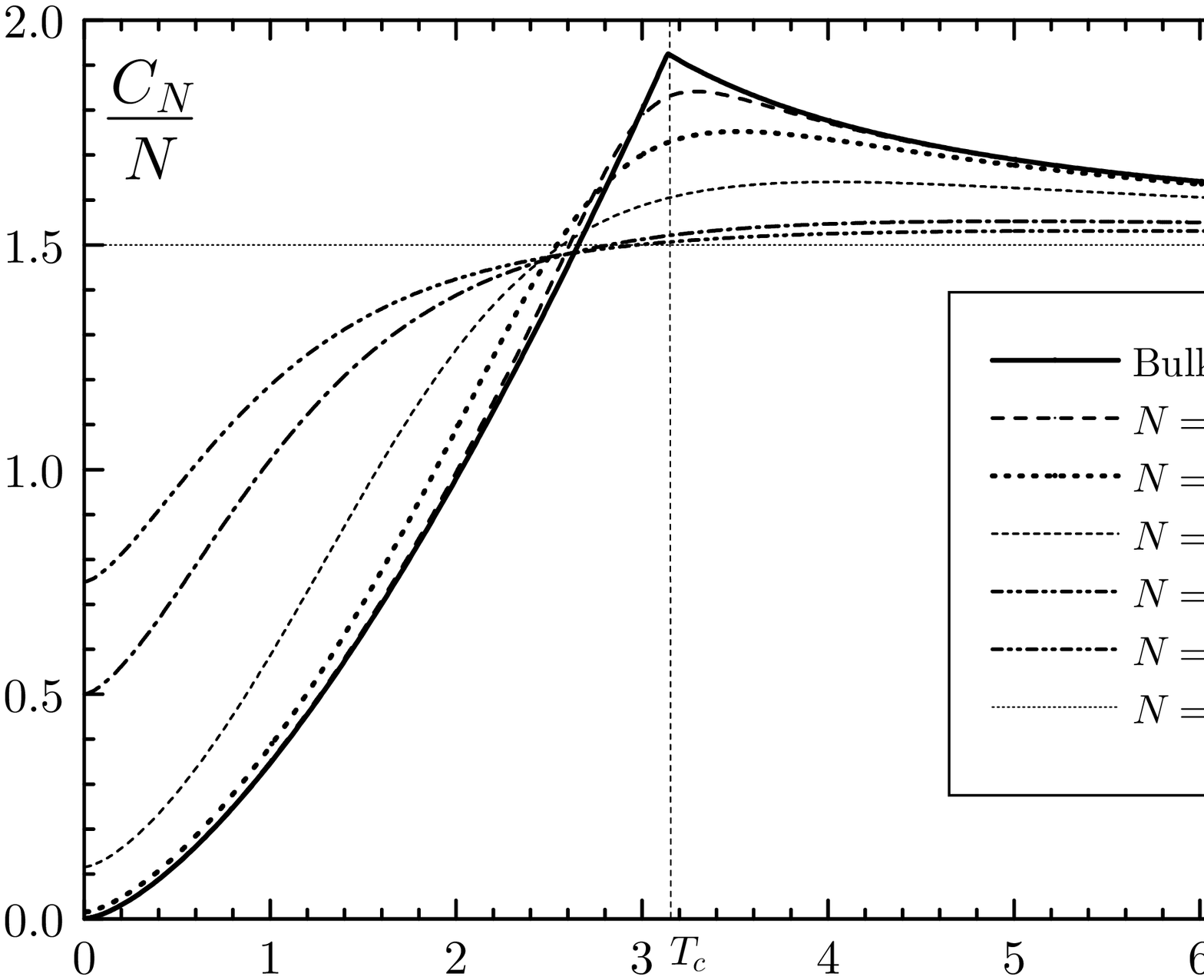}}
\bigskip
\caption{
Heat capacities in the ideal case.
}
\end{figure}

\bigskip
\begin{figure}
\epsfxsize=85mm
\centerline{\epsfbox{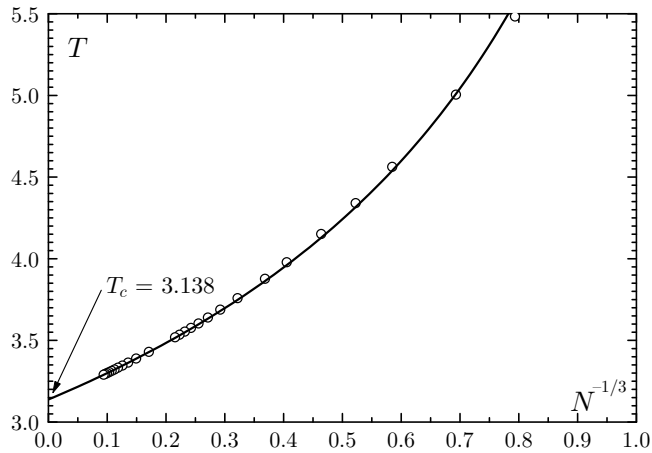}}
\bigskip
\caption{
Temperature of the $C_N/N$ maxima in the ideal case,\\
Solid line shows dependence (\protect\ref{extr1}),\\
circles correspond to the data from Table~I.}
\end{figure}
\bigskip

The results are also given in Table~I.

\bigskip
\parbox[b]{\textwidth}
{
\begin{center}
{
\begin{tabular}{|c|c|c||c|c|c|}
\hline\hline
{\vrule height1.8em width0em depth1.0em}
$N$ &$\rho\left[\lambda_c^{(N)}\right]^3$&$ C_N^{\max}/N$&$N$&$\rho
\left[\lambda_c^{(N)}\right]^3$&$C_N^{\max}/N$\\\hline\hline
  1 & --- &1.500&  90 &2.185  &1.747\\\hline
  2 &1.131&1.532& 100 &2.200  &1.753\\\hline
  3 &1.297&1.553& 200 &2.286  &1.783\\\hline
  5 &1.490&1.582& 300 &2.328  &1.799\\\hline
  7 &1.606&1.602& 400 &2.354  &1.809\\\hline
 10 &1.717&1.624& 500 &2.373  &1.816\\\hline
 15 &1.830&1.649& 600 &2.387  &1.822\\\hline
 20 &1.902&1.666& 700 &2.399  &1.827\\\hline
 30 &1.993&1.690& 800 &2.408  &1.831\\\hline
 40 &2.050&1.706& 900 &2.416  &1.834\\\hline
 50 &2.091&1.718&1000 &2.423  &1.837\\\hline
 60 &2.123&1.727&1100 &2.429  &1.839\\\hline
 70 &2.147&1.735&1200 &2.434  &1.842\\\hline
 80 &2.168&1.742&$\infty$&2.612&1.926\\\hline\hline
\end{tabular}

\bigskip
{TABLE I. Main results for the ideal systems}
}
\end{center}
}

As one can see from Fig.~1, good qualitative and satisfactory quantitative
agreement is reached at $N=100$. Although, typical elements of the heat 
capacity curve appear at smaller $N$, e.~g., a clearly defined maximum
is found at $N=10$--$15$. As we shall show further, these very values
might be considered as the limit between ``individual'' and statistic
behaviour of the system with $N$ particles.

\section{Interaction in the hard-spheres potential
approximation}
In a many-body system, it is a very complicated problem to take into account 
the interaction and it has not as yet been completely solved.
Specifically, the systems with a small number of particles give a possibility
to consider this problem from the first principles, i.~e., to include
the interaction into the calculations explicitly.
From the mathematical point of view, the computations become heavily
complicated while the number of particles increases, and the direct
interaction account soon
turns out to be impossible.

We will proceed from the expression for the density matrix $R_N$ in the free particles case.
In the systems like \He4 small interatomic distances are essential
when atoms behave almost as hard spheres.
Thus, the wavefunction should vanish rapidly as soon as 
the distance between any two atoms approaches the hard sphere diameter.

Due to this we shall take the interaction into account by the substitution:
of $\Delta_N$ in (\ref{Z_int}) with $\Delta_N P_N$ where $P_N$
is a function of the particles coordinates:
\be \label{Z_intP}
Z_N=\frac1{N!}\frac1{\lambda^{3N}}\int d\r_1\dots\int d\r_N
\hskip3pt\Delta_N P_N.
\ee

For the sake of simplicity we suppose the interaction to have a pair character
since the obtained results are expected to be rather qualitative than
quantitative.
In the case of hard spheres potential we have
\be
&&P_N(\r_1,\dots,\r_N)=\prod_{i=1}^{N-1}\prod_{j=2}^N P(i,j),\nonumber\\
&&P(i,j)=P(|\r_i-\r_j|)=P(r_{ij}),\nonumber\\
&&P(i,j)=1+f(i,j)=\left\{
\begin{array}{cc}
0, & r_{ij}\le a \\
1, & r_{ij}>a
\end{array}
\right.,
\ee
$a$ is the hard sphere diameter.

One can write the partition function as
\be
Z_N=\frac1{\rho\lambda^3}\sum_{l=1}^N b'_l(N)Z_{N-l},
\qquad Z_0\equiv1,
\ee

\be
\lambda=\left( {2\pi\beta \hbar^2} \over {m^*} \right)^{1/2},
\ee
where $m^*$ is the effective mass of \He4 atom, $m^*\simeq1.7m$~\cite{Visn1}.

We integrate (\ref{Z_intP}) in two steps. The first one is
{\it the circular approximation (CA)} and the second one is
{\it the free-volume approximation (FVA)}.

1)~CA is to the effect that
only the interparticle interactions in the circle
1---2---3---$\dots$---$N$---1 are taken into account. It means we neglect
the interaction between the first and the third, fourth, etc. particles,
between the second and fourth, fifth, etc. particles, 
\ldots$\;$.
This makes it possible to integrate (\ref{Z_intP}) using Fourier transformation
as we did in (\ref{Z_int}).
Using also FVA we obtain for the coefficients
$b'_l(N)$:
\be \label{bprime}
b'_l(N)=
\left\{
\begin{array}{cc}
\left[1-\frac{\vo}{V}(N-l)\right]^2 \left[1-\frac{\vo}{V}(N-3)\right]
\prod\limits_{k=3}^{l-1}\left[1-\frac{\vo}{V}(N-k)\right]
I_{l-1},
& l>3 \\
{\vrule height2em width0em}
\left[1-\frac{\vo}{V}(N-l)\right]^l I_{l-1}, & l\le 3
\end{array}
\right.
\ee
Here we have used such designations:
\be
\vo=\frac{4}{3}\pi a^3,
\ee

\be
I_n=\left\{
\begin{array}{cc}
8(4\pi)^{n-1}
\begin{displaystyle}
\int_0^{\infty}dQ\hskip 5pt Q^2\left(\frac{1}{4\pi}e^{-Q^2/4\pi}
-\frac{1}{Q}\int_0^{a/\lambda}x\; \sin Qx \; e^{-\pi x^2}\;dx\right)^{n+1}
\end{displaystyle}
, & n>0 \\
{\vrule height2em width0em}
1, & n=0
\end{array}
\right.
\ee

2)~In FVA we indirectly take into account the interactions neglected 
in CA.
For this purpose
each integral over $\r_i$, $i=1,\dots,l$
is multiplied by the factor $(1-n\vo/V)$ where $n$ equals to the number of
multipliers $P(i,j)$ which are not taken into account in the Fourier
transformation. To clarify the above statement we consider the most simple
non-trivial example with $N=4$. According to (\ref{bprime}) we have:
\begin{list}{$\bullet$}{}
{
\item{$b'_1(4)=\left[1-\displaystyle{{3\vo\over V}}\right] I_0$.\ \
Integral over $\r_1$ ``hooks'' the folloving {\it three} multipliers:
$P(1,2)$, $P(1,3)$, $P(1,4)$.}
\item{$b'_2(4)=\left[1-\displaystyle{{2\vo\over V}}\right]^2 I_1$.\ \
Integral over $\r_1$ ``hooks'' $P(1,2)$, $P(1,3)$, $P(1,4)$ but $P(1,2)$
is included into $I_1$. Integral over $\r_2$ ``hooks'' more $P(2,3)$ and
$P(2,4)$.}
\item{$b'_3(4)=\left[1-\displaystyle{{\vphantom{2}\vo\over V}}\right]^3 I_2$.\ \
Each integral over $\r_i$ ``hooks'' additional $P(i,2)$.}
\item{$b'_4(4)=\left[1-\displaystyle{{\vphantom{2}\vo\over V}}\right]^2 I_3$.\ \
$P(1,3)$ and $P(2,4)$ are not included into $I_3$.}
}
\end{list}

Thus, FVA improves the circular approximation.
But it is not possible to use it for large $l$ and $N$ because in this
case a big number of $P(i,j)$ multipliers will not be integrated in the
proper way but via formal factors.

In the calculations we assume the hard sphere
diameter $a=2.1$~\AA. The heat capacity curves exhibit a fast
approaching of the maximum point to a value from the region
of 2.2--2.3~Š while $N$ is increasing.
The exact $\lambda$-transition temperature is 2.17~K.

As has already been mentioned, in the ideal case the linear (over $N^{-\varepsilon}$)
dependence
fits the sequence of $C_N/N$ maxima quite well
\be
  { C_N^{\max}\over N}\simeq C^{\max}_{\infty} - a N^{-\varepsilon},
  \quad \varepsilon=1/3.
\ee
Assuming the same form for the hard spheres system one can find
$C^{\max}_{\infty}=13.6$
and $\varepsilon=0.0608$ making the maxima sequence as follows:
\be \label{Cn1}
  { C_N^{\max}\over N}\simeq 13.6 - 12.9\, N^{-\varepsilon},
  \quad \varepsilon=0.0608.
\ee

Although small $N$ values of 12, 13 do not allow one to identify
the behaviour at $N\to\infty$ we obtain the values of heat capacity
maximum from the range of experimental data~\cite{NIST,HeliumConfined,Lipa96}.

\bigskip\bigskip
\begin{figure}
\epsfxsize=85mm
\centerline{\epsfbox{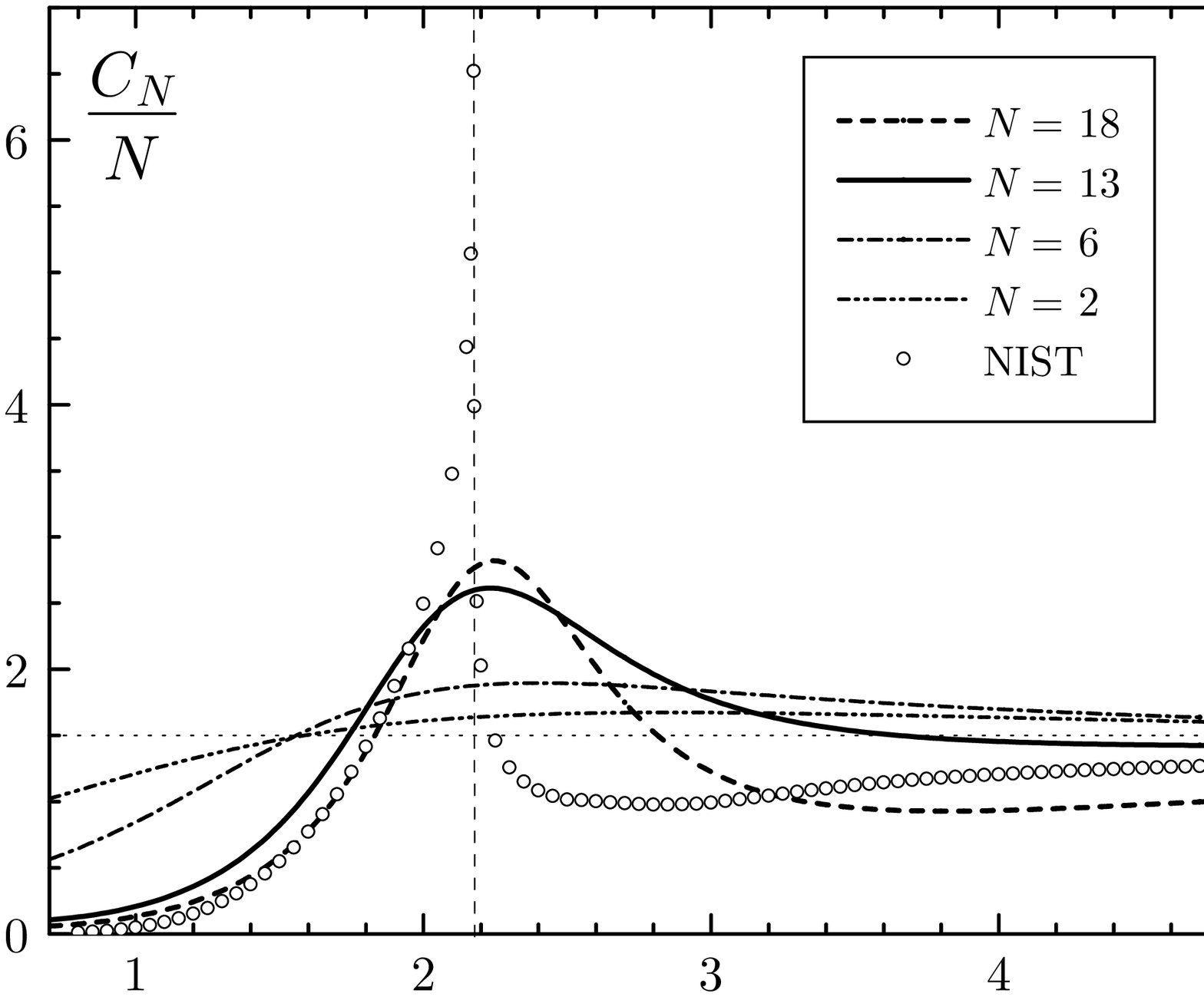}}
\bigskip
\caption{
Heat capacities in the free volume and circular approximation.
The circles show the NIST $C_V$ data~\protect\cite{NIST}.
}
\end{figure}
\bigskip

One can write the exact expression for the partition function
similiar to that in the case of CA and FVA:
\be
Z_N=\frac1{\rho\lambda^3}\sum_{l=1}^N b^*_l(N)Z_{N-l},
\qquad Z_0\equiv1.
\ee
From expansion (\ref{Delta_expan}) and formula (\ref{Z_intP}) we obtain
(without approximations):
\be
b^*_l(N)&=&\frac1V\frac1{(N-l)!}\frac1{(\lambda^3)^{N-1}}\int d\r_1\dots
\int d\r_l\hskip5pt K_{12}\dots K_{l-1,l}K_{l1} \nonumber \\
&&{}\times \prod^{l-1}_{i=1}\prod^l_{j=i+1}
P(i,j)\hskip5pt\B_{lN}(1,\dots,l),
\ee

\be
\B_{lN}(1,\dots,\l)=\int d\r_{l+1}\dots\int d\r_N\frac
{\Delta^{(1,\dots,\l)}_{N-l}P^{(1,\dots,\l)}_{N-l}
\prod\limits^l_{i=1}\prod\limits^N_{j=l+1}
P(i,j)} {Z_{N-l}},
\ee

\be
P^{(1,\dots,\l)}_{N-l}=\prod^{N-1}_{i=l+1}\prod^N_{j=i+1}P(i,j).
\ee

We also introduce the following designations
to simplify further records:
\be
R(1,2)=\int d\r_3\hskip5pt P(1,3)P(2,3),
\ee

\be
\widetilde K_{ij}=K_{ij}P(i,j).
\ee

One can write the coefficients $b^*_l(N)$ independently of the
interatomic potential, i.~e. for any $P_N$:
\be
b^*_1(1)&=&1, \\
b^*_1(2)&=&\frac1{V^2}\int d\r_1\int d\r_2 \hskip5pt P(1,2), \\
b^*_1(3)&=&\frac{\displaystyle{\int d\r_1\int d\r_2\hskip3pt
         P(1,2)R(1,2)\left(1+K_{12}K_{21}\right)}}
         {\displaystyle{\int d\r_1\int d\r_2\hskip3pt
         P(1,2)\left(1+K_{12}K_{21}\right)}},\\
b^*_2(2)&=&\frac1V\frac1{\lambda^3}\int d\r_1\int d\r_2\hskip3pt
         K_{12}K_{21}P(1,2),\\
b^*_2(3)&=&\frac1{V^2}\frac1{\lambda^3}\int d\r_1\int d\r_2\hskip3pt
         K_{12}K_{21}P(1,2)R(1,2),\\
b^*_3(3)&=&\frac1V\frac1{(\lambda^3)^2}\int d\r_1\int d\r_2\int d\r_3\hskip3pt
         \widetilde K_{12}\widetilde K_{23}\widetilde K_{31}.
\ee

These expressions might be written via one-dimensional integrals:
\be
b^*_1(2)&=&\frac{4\pi}V \int^{\infty}_0 P(r_{12}) r_{12}^2\hskip3pt dr_{12},\\
b^*_2(2)&=&\frac{4\pi}{\lambda^3}\int^{\infty}_0 K_{12}^2P(r_{12})r_{12}^2
         \hskip3pt dr_{12},\\
R(r_{12})&=&\frac8{r_{12}}\int^{\infty}_0\frac{dq}q\sin qr_{12}\left\lbrace
          \int^{\infty}_0 rP(r)\sin qr\hskip3pt dr\right\rbrace^2,\\
b^*_3(3)&=&\frac{32\pi}{(\lambda^3)^2}\int^{\infty}_0\frac{dq}q
      \left\lbrace \int^{\infty}_0 K_{12}P(r_{12})r_{12}\sin qr_{12}\hskip3pt
      dr_{12}\right\rbrace^3,\\
b^*_2(3)&=&\frac{4\pi}{V\lambda^3}\int^{\infty}_0K_{12}^2
         P(r_{12})R(r_{12})r_{12}^2\hskip5pt dr_{12}.
\ee
In the hard spheres approximation integration is made simply. We obtain
\be
b^*_1(1)&=&1,\\
b^*_1(2)&=&1-\frac{\vo}V,\\
b^*_2(2)&=&\frac1{2^{3/2}}
         \left\lbrace1-\erf\left(\all\sqpi\right)\right\rbrace
         +\all e^{-2\pi\als2},\\
b^*_1(3)&=&\frac{V-3\vo+\frac{81}{32}\frac{\vo^2}V+\lambda^3b^*_2(3)}
         {V b^*_1(2)+\lambda^3b^*_2(2)},\\
b^*_2(3)&=&\frac{V-\vo}V\left\lbrace\all\left(e^{-2\pi\als2}-2e^{-8\pi\als2}
         \right)+\frac1{2^{3/2}}\left(\erf\left(2\all\sqpi\right)
         -\erf\left(\all\sqpi\right)\right)\right\rbrace \\
&&{}+\frac{\pi a^2\lambda}V\left\lbrace\al2\left(e^{-2\pi\als2}-
    4e^{-8\pi\als2}\right)
    +\frac1{2\pi}\left(e^{-2\pi\als2}-e^{-8\pi\als2}\right)\right\rbrace \\
&&{}+\frac{\pi\lambda^3}{12V}\left\lbrace e^{-2\pi\als2}
    \left(\al4+\frac1{\pi}\al2+\frac1{2\pi^2}\right)
    -e^{-8\pi\als2}\left(16\al4+\frac4{\pi}\al2+\frac1{2\pi^2}
    \right)\right\rbrace \\
&&{}+\frac{V-2\vo}V\left\lbrace\frac1{2^{3/2}}
    \left(1-\erf\left(2\all\sqpi\right)\right)
    +2\all e^{-8\pi\als2}\right\rbrace,\\
b^*_3(3)&=&32\pi\int^{\infty}_0\frac{dQ}Q
\left\lbrace \frac{Q}{4\pi}e^{-Q^2/4\pi}-\int^{\alls}_0
x\;\sin Qx\;e^{-\pi x^2}\; dx
\right\rbrace^3.
\ee

Our $R$ coincides with the correspondent function from~\cite{Nijboer}:
\be
R(r_{12})=\int d\r_3\hskip5pt (1+f_{13})(1+f_{23})=V-2\vo+g_1(r_{12}),
\ee
The function $g_1(r_{12})$ is given by
\be
g_1(r)=\int d\r_3\hskip3pt f_{13}f_{23}=
\left\{
\begin{array}{cc}
\frac23\pi a^3\left(2-\frac32\frac{r}{a}+
\frac18\left(\frac{r}{a}\right)^3\right), & r<2a \\
0, & r\ge 2a
\end{array}
\right.
\ee

Meanwhile we cannot write exact expressions for greater $N$.
The comparison of the exact results and those in
CA and FVA is presented in Table~II.

\bigskip
\parbox{\textwidth}
{
\begin{center}
\begin{tabular}{|r|c|c|c|c|c|c|}
\hline\hline
&&\multicolumn{2}{c|}
{{\vrule height1.5em width0em depth0.8em}
Approximation}
&&\multicolumn{2}{c|}{Exact values}\\
\cline{3-4}\cline{6-7}
\raisebox{6pt}[0pt][0pt]{$N\:$}
 &{}&{\vrule height 14pt width 0pt depth 6pt}
 $ T_c^{(N)}$ & $C_N^{\max}/N$ && $T_c^{(N)}$ & $C_N^{\max}/N$\\
\cline{1-1}\cline{3-4}\cline{6-7}
 1\phantom{$^*$} &{}&  --- & 1.50 &&  --- & 1.50 \\
\cline{1-1}\cline{3-4}\cline{6-7}
 2\phantom{$^*$} &{}& 2.86 & 1.67 && 2.86 & 1.67 \\
\cline{1-1}\cline{3-4}\cline{6-7}
 3\phantom{$^*$} &{}& 2.54 & 1.78 && 2.19 & 1.83 \\
\cline{1-1}\cline{3-4}\cline{6-7}
 4\phantom{$^*$} &{}& 2.28 & 1.90 &&  --- &  --- \\
\cline{1-1}\cline{3-4}\cline{6-7}
 5\phantom{$^*$} &{}& 2.25 & 2.01 &&  --- &  --- \\
\cline{1-1}\cline{3-4}\cline{6-7}
 6\phantom{$^*$} &{}& 2.24 & 2.11 &&  --- &  --- \\
\cline{1-1}\cline{3-4}\cline{6-7}
 7\phantom{$^*$} &{}& 2.23 & 2.20 &&  --- &  --- \\
\cline{1-1}\cline{3-4}\cline{6-7}
 8\phantom{$^*$} &{}& 2.23 & 2.29 &&  --- &  --- \\
\cline{1-1}\cline{3-4}\cline{6-7}
 9\phantom{$^*$} &{}& 2.23 & 2.37 &&  --- &  --- \\
\cline{1-1}\cline{3-4}\cline{6-7}
10\phantom{$^*$} &{}& 2.23 & 2.44 &&  --- &  --- \\
\cline{1-1}\cline{3-4}\cline{6-7}
11\phantom{$^*$} &{}& 2.23 & 2.50 &&  --- &  --- \\
\cline{1-1}\cline{3-4}\cline{6-7}
12\phantom{$^*$} &{}& 2.23 & 2.56 &&  --- &  --- \\
\cline{1-1}\cline{3-4}\cline{6-7}
13$^*$ &{}& 2.23 & 2.61 &&  --- &  --- \\
\cline{1-1}\cline{3-4}\cline{6-7}
18\phantom{$^*$} &{}& 2.23 & 2.82 &&  --- &  --- \\
\cline{1-1}\cline{3-4}\cline{6-7}
{\vrule height1.0em width0em}
$\infty$\phantom{$^*$} &{}& 2.2--2.3 & $13.6^{**}$ && --- & --- \\
\hline\hline
\end{tabular}

\medskip
\begin{list}{x}{}
\setlength{\leftmargin}{5cm}
\setlength{\rightmargin}{5cm}
\setlength{\labelwidth}{50pt}
\item[$^{*}$]
{We suppose that the system of 12--13
particles behaves almost as a bulk since the atoms in \He4 are packed
compactly enough to say that one atom is surrounded by 12 atoms.
Therefore, 13 atoms constitute the first closed sphere (see Fig.~5).}
\item[$^{**}$]{The experimental value is $\simeq16$.
The heat capacity in the $\lambda$-point was believed to
be infinite for a long time but recent experiments show the finite
peak~\cite{Lipa96}.}
\end{list}
\end{center}

\begin{center}
\bigskip
{TABLE II. Results for the interacting systems}
\end{center}
}

\begin{figure}
\epsfxsize=85mm
\centerline{\epsfbox{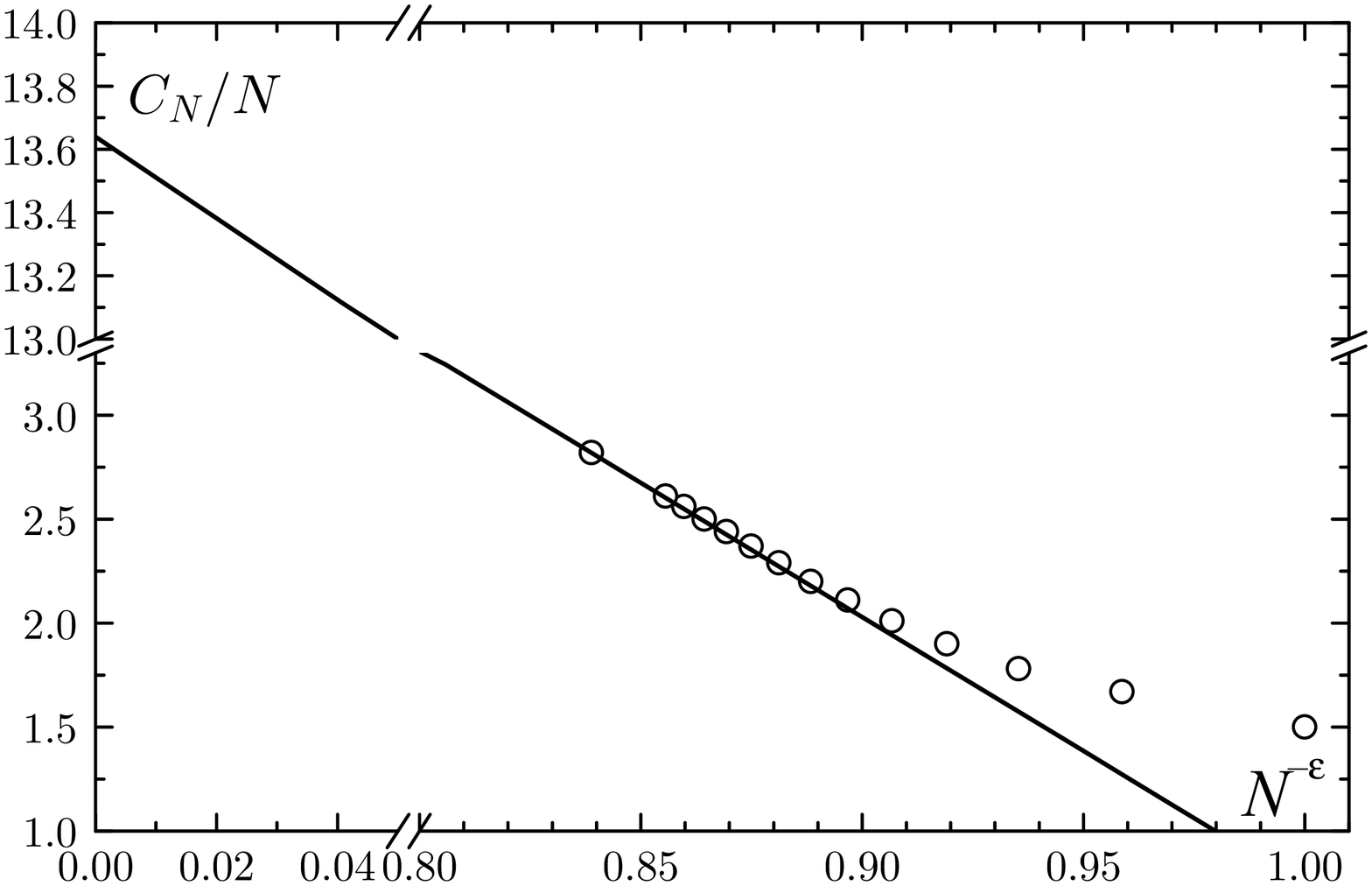}}
\bigskip
\caption{
$C_N^{\max}/N$ values for the hard spheres system.\\
Solid line shows dependence (\protect\ref{Cn1}),\\
circles correspond to the data from Table~II.
}
\end{figure}
\bigskip

\begin{figure}
\unitlength=1mm
\begin{picture}(75,60)
\put(0,15){
\epsfxsize=40mm
\epsffile[138 262 456 580]{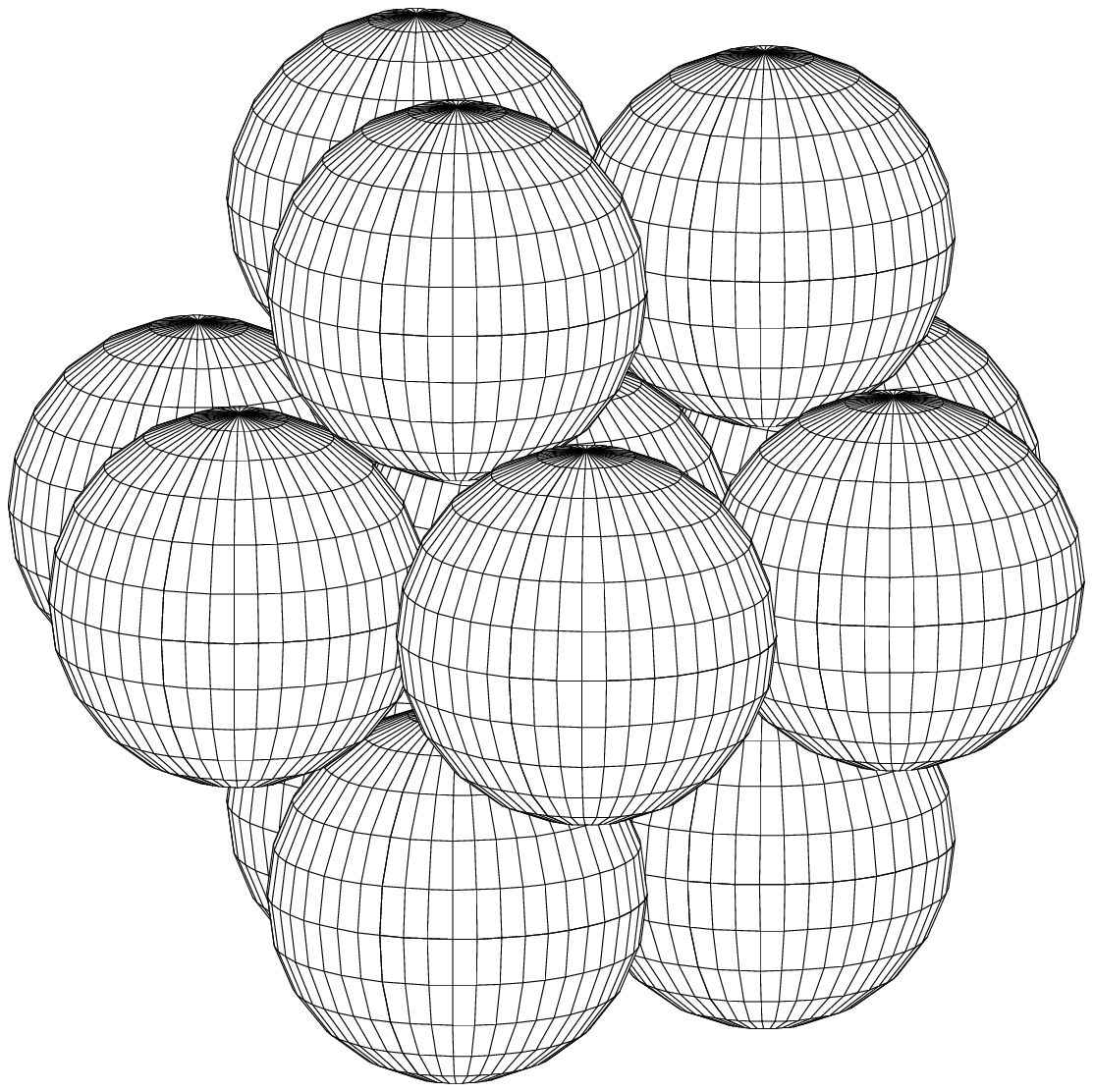}}
\epsfxsize=35mm
\put(45,20){\epsffile[157 609 244 679]{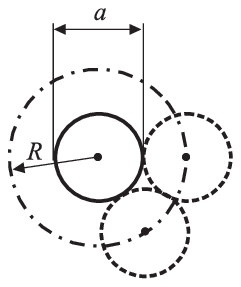}}
\put(20,10){(a)}
\put(55,10){(b)}
\end{picture}
\caption{Helium-4 atoms as hard spheres.\newline
(a)~The first closed sphere of 12 atoms surrounding 1 central atom.
One atom is not seen since it is situated exactly behind the central
atom.\newline
(b)~Solid line shows the ``surface of atom'',
dashed line shows the ``surface of alien atoms'',
dashed-dotted line shows ``surface of inaccessible space''.
Thus, assuming the volume of ``inaccessible space'' $4\pi R^3/3$ to be
the volume occupied
by 1 atom one can obtain the hard sphere diameter $a\simeq2.2$~\AA.
}
\end{figure}

Our expressions for the partition function  
have the correct form $Z^{\rm (cl)}_N$ in the classical limit 
($\hbar\to 0$)~\cite{Isihara}.
\be
Z^{\rm (cl)}_N=\frac{1}{N!}\frac{1}{\lambda^{3N/2}}\int d\r_1\dots\int d\r_N
\hskip5pt \prod_{1\le i<j\le N} P(i,j).
\ee
It means that the proposed method is applicable for a wide temperature
range

\section*{\bf Conclusion}
It is shown that proposed method leads to a good agreement between
the calculated results and the experimental data. We obtained
not only qualitive but even satisfactory quantitive fit as well.

The expansions over $1/N$ obtained in this work demonstrate
the existence of some parameter
$\varepsilon$ having the value of $1/3$ for the ideal system. We estimated
$\varepsilon=0.0608$ from the numerical analysis of the results.
The theoretical calculation of this parameter will be our next problem
along this line of research.

\section*{ACKNOWLEDGEMENTS}
The authors appreciate the help of Dr.~Daniel~G.~Friend
with the helium-4 thermophysical data.

\end{document}